\documentclass[aps,twocolumn,superscriptaddress,nofootinbib]{revtex4}

\usepackage{exscale}
\usepackage{graphicx}
\usepackage{amsmath}
\usepackage{latexsym}
\usepackage{amsfonts}
\usepackage{amssymb}\begin{document}

\title{Trapped ion chain as a neural network}

\author{M. Pons}
\affiliation{Departamento de F\'isica Aplicada I, Universidad del Pa\'is Vasco, E-20600 Eibar, Spain}
\author {V. Ahufinger}
\affiliation{ICREA and Grup d'\`Optica, Departament de F\'isica, Universitat Aut\`onoma de Barcelona, E-08193 Bellaterra, Barcelona, Spain.}
\author {C. Wunderlich}
\affiliation{Fachbereich Physik, Universit\"at Siegen, 57068 Siegen, Germany.}
\author {A. Sanpera}
\affiliation{ICREA and Grup de F\'isica Te\`orica, Departament de F\'isica, Universitat Aut\`onoma de Barcelona, E-08193 Bellaterra, Barcelona, Spain.}
\author{M. Lewenstein}
\affiliation{ICREA and Institut de Ci\`encies Fot\`oniques, 
E-08034 Barcelona, Spain.}



\begin{abstract}
We demonstrate the possibility  of realizing a neural network in a chain
of trapped ions with induced long range interactions. 
Such models permit to store information distributed over the whole system.
The storage capacity of such network, which depends on the phonon spectrum of the system, can
be controlled by changing the external trapping potential and/or by applying longitudinal local magnetic fields. The system properties suggest the possibility of  implementing
{\sl robust distributed} realizations of quantum logic.

\end{abstract}
\maketitle

The Cirac-Zoller trapped ion computer \cite{cirac} has become one of
the paradigmatic models to implement a quantum
computer \cite{general}. Recently, spectacular experimental progress
in both realization of simple algorithms and implementation of
quantum logic has been achieved using a few ions qubits (c.f.
\cite{experiments}).

Still, serious experimental challenges remain on the way to a
large-scale ion trap quantum computer. One should stress that while
the achievement of an all-purpose quantum computer in the next years
seems to be difficult, one can be quite optimistic about the
applications of chains of trapped ions as quantum simulators.
Recently, it has been shown that long range pairwise interaction
between individual spins is induced in an ion trap when introducing
an additional state-dependent force acting on the ions
\cite{christoph,Wunderlich02}. The properties of such ion spin
molecules can be designed by variation of the parameters characterizing the
ion trap. Later, a state-dependent optical force has also been shown to
evoke long range couplings and has been 
proposed to simulate general Heisenberg interactions and to 
investigate quantum phase transitions in ion traps \cite{porras}.
Here we show that ion spin systems can serve to realize a neural
network (NN) model. Neural networks are a prototype model of
parallel distributed memory \cite{Amit,parisi}, and
have been intensively studied by physicists since the famous paper
by Hopfield \cite{hopfield}. These disordered systems with long
range interactions typically present a large number of metastable
(free) energy minima, like in spin glasses \cite{parisi}.
These states can be used to store information distributed over the
whole system. The patterns stored by the network have large basins
of attraction in the thermodynamical sense which means that even
fuzzy patterns are recognized as perfect ones. For this reason,
attractive NN's can be used as associative memory. At the same time,
NN are robust, so that destroying even a large part of the network
does not necessarily diminish the network performance.
The above listed properties
make NN's interesting for {\it distributed} 
quantum information storing and processing where a quantum bit 
would correspond to two collective modes of the ion's 
chain instead of two internal states of each individual ion.
Different approaches to exploit the potential use of NN models 
for quantum information processing have been already 
discussed in the literature \cite{nn,sen}. In \cite{sen} we have shown  
that it is possible to create next neighbour entanglement in an ideal
NN model. Here we propose, for the first time, 
a feasible implementation of NN using a chain of trapped ions. 

Before explaining the details of NN implementation we first remind
the readers the main features of the Hopfield model
\cite{hopfield}, and discuss the effective Hamiltonian derived in
Ref.\cite{Wunderlich02,porras}. The similarities between both models
suggest the possibility of using a chain of trapped ions as a NN. 
To demonstrate this assumption we find 
the storage capacity of the system and its robustness which are
the most appealing features of NN for distributed quantum information.
Thus, the question of ergodicity and, therefore, 
the ability of the system to perform an associative memory is not
considered here. 
We show that the storage capacity, which is determined by the phonon 
spectrum of the system, can be controlled by modifying the
shape of the external axial trapping potential and/or by applying
longitudinal magnetic fields. Although the storage capacity is a
classical property of the network, spin ion systems permit also 
to study quantum neural networks by using a transverse magnetic field or an
optical field that effectively simulates an additional transverse
magnetic field. Transverse magnetic fields should permit tunneling
processes between stored patterns and to realize for example of
quantum stimulated annealing \cite{parisi,ags}. 
This study is beyond the scope
of this letter and will be treated elsewhere.

Following the models of Hopfield \cite{hopfield} and Little
\cite{little}, a neuron can be viewed as an Ising spin with two
possible states: an "up" position $(S=+1)$ and a "down" position
$(S=-1)$ depending on whether the neuron has or has not fired an
electromagnetic signal, in a given interval of time
\cite{excellent}. The state of the network of $N$ neurons at a
certain time is defined by the instantaneous configuration of all
the spin variables $\lbrace S_i\rbrace$ at this time. The dynamic
evolution of these states is determined by the interactions among
neurons, $J_{ij}$, which are symmetric, so, for any pair
of neurons, $J_{ij}=J_{ji}$. Also full connectivity is assumed,
that is, every neuron can receive an input from any other one, and
can send an output to it. The Hamiltonian of the system can be
written as:
\begin{equation}
H=-\frac{1}{2}\sum_{i,j}^{N}J_{ij}S_i S_j + h\sum_{i}^{N}S_i,
\label{ham_neural}
\end{equation}
where $h$ corresponds to an external magnetic field.
The interactions are determined by
the patterns or configurations of spins to be stored in the network.
These patterns will be learned if the system is able to accommodate
them as attractors, implying that a large set of initial 
configurations of the network will be driven dynamically (after sufficient
enough time) to those patterns. 
A possible choice of the interactions is:
\begin{equation}
J_{ij}=\frac{1}{N}\sum _{\mu=1}^p \xi_i^{\mu}\xi_j^{\mu}\,,
\label{int_neural}
\end{equation}
with $i\not= j$. The $p$ sets of $\lbrace \xi_{i}^{\mu}\rbrace = \pm 1$ are the patterns that we wish to store. 
The network will have the capacity of storage and retrieval of information if the dynamical stable configurations
(local minima) reached by the system $\lbrace S_{i} \rbrace$ are correlated with the learned ones $\lbrace \xi_{i}^{\mu}\rbrace$. Despite the fact that the interactions have been constructed to guarantee that certain specified patterns be fixed points of the dynamics, the non-linearity of the dynamical process may induce additional attractors, the so called spurious states.

Recently it has been shown that the Hamiltonian describing a linear
chain of harmonically trapped ions exposed to a magnetic field
gradient \cite{Wunderlich02} or interacting with convenient laser fields \cite{porras} can be transformed into an effective spin-spin Hamiltonian
with long range interactions mediated by the collective motion of
the ions:

\begin{equation}
H=-\frac{1}{2}\sum_{\alpha,i,j}J^{\alpha}_{ij}\sigma^{\alpha}_i\sigma^{\alpha}_j+\sum_{\alpha,i}B^{\alpha}_i\sigma^{\alpha}_i\,,
\label{hamiltonian}
\end{equation}
where:
\begin{equation}
J^{\alpha}_{ij}=\frac{(F^{\alpha}_{ ij})^2}{m}\sum_n
\frac{M^{\alpha}_{i,n}M^{\alpha}_{j,n}}{\omega^2_{\alpha,n}}\,,
\label{int}
\end{equation}
with $\alpha= x, y $ and $z$ being the spatial directions, $i, j$
labeling the ions, $\sigma$ being the Pauli matrices, $F^{\alpha}_{ij}$ 
the forces in the $\alpha$ direction, $m$ the mass of the ion, 
$\omega_{\alpha,n}$ the angular frequency of vibrational mode $n$.
$M^{\alpha}_{i,n}$ are the
unitary matrices that diagonalize the vibrational Hamiltonian:
\begin{equation}
M^{\alpha}_{i,n}{\kappa}^{\alpha}_{i,j}
M^{\alpha}_{j,m}=\omega^2_{\alpha,n} \delta_{nm}\,,
\label{ms}
\end{equation}
where ${\kappa}^{\alpha}_{i,j}$ are the elastic constants of the
chain \cite{goldstein}. The coefficient $M^{\alpha}_{i,n}$ gives the
scaled amplitude of the local oscillations of ion $i$ around its
equilibrium position when the collective vibrational mode $n$ is
excited. Thus, the eigenvectors of $M$ describe each ion's
contribution to a given vibrational mode while the eigenvalues
provide the oscillation frequencies, $\omega_{\alpha,n}$ of the
collective modes.

\begin{figure}
\includegraphics[width=1.0\linewidth]{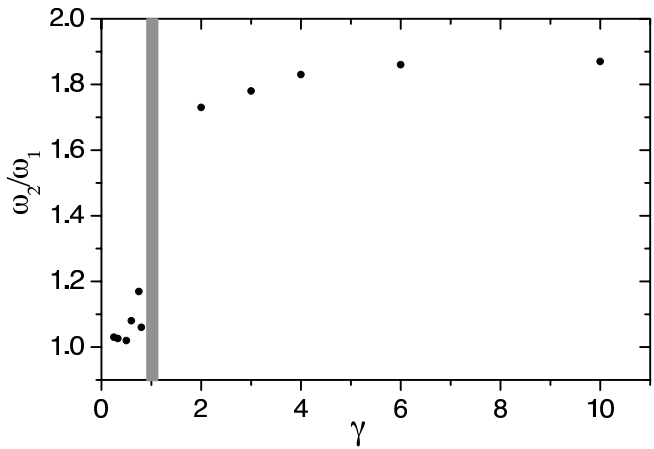}
\caption{Ratio between the frequencies of the second and the first
vibrational modes as a function of the exponent of the trapping
potential, $\gamma$} \label{om1om2b}
\end{figure}

The external trapping frequencies are chosen such that the laser
cooled ions crystallize in a linear chain ({\it i.e.},
$\Omega_{x}=\Omega_{y}\gg \Omega_{z}$) and the external 
forces are such that on the $z-$axis
$F^{x}_{ij}=F^{y}_{ij}=0$ holds, and the index $\alpha$ is
dropped. From now on we will consider the case
of a chain of ions interacting with external laser fields \cite{porras} which
allow us to consider site independent forces and magnetic field, 
$F_{ij}=F$, $B'_{i}=B'$ \cite{noteta}. 
Thus, Eqs.(\ref{ham_neural}) and
(\ref{hamiltonian}) have the same form and the possibility to
implement a classical neural network using a linear chain of ions
arises. We substitute the Pauli matrices in Eq.(\ref{hamiltonian}) 
by Ising spins $S=\pm 1$, where the effective spin corresponds to the 
internal state of the ion.
Nevertheless, there are some important differences
between both models. First, in the Hopfield model, the interactions
given by Eq.(\ref{int_neural}) are determined by the patterns to be
stored $\lbrace \xi_{i}^{\mu}\rbrace=\pm 1$ while in the trapped ion
chain, the interactions are fixed by the
coefficients $M_{i,n}$ that do not necessarily equal $\pm 1$. Second,
in Eq.(\ref {int_neural}), $p$ corresponds to the number of
patterns to be stored, which in the limit of large number of spins
$N$ is bounded from above by $p=0.14N$  \cite{Amit}. In our model,
the sum (Eq.(\ref{int})) extends to the total number of
vibrational modes which is larger than the total number of
stored patterns. And finally, in the Hopfield model all the
patterns have the same weight while in the ion chain each
vibrational mode (Eq.(\ref{int})) is weighted 
by  $1/ \omega^{2}_{n}$. Due to this fact, 
to reproduce as close as possible a NN model, 
the vibrational modes should be ideally
almost degenerate in energy. Moreover, the corresponding patterns 
must have large basins
of attraction, i.e. they should correspond to sufficiently
different configurations of spins so that each configuration is
dynamically recovered even if several spins are randomly flipped.
These requirements are not achievable with an harmonic trapping
potential due to the fixed frequency ratio between the two lowest
vibrational modes \cite{james}. The desired degeneracy can be achieved 
either by engineering an adequate external magnetic field or by 
using a trapping potential of the form of  
$\mid x \mid^{\gamma}$ with $\gamma \le 0.8$ (see Fig.\ref{om1om2b}).

To check the feasibility of implementing a NN model in these ion spin
systems,  
we calculate first the phonon spectrum, 
impose the learning rule (Eq.(\ref{int})),
and probe the correlation between the dynamically stable
configurations and the stored ones. Therefore, 
the first step consists in finding the
equilibrium positions of the ions 
and calculate the vibrational modes and frequencies 
by using a standard diagonalization procedure.
Once the complete vibrational spectrum of the system is known, we
calculate the interactions among spins $J_{ij}$,
and evaluate the energy of the system. The dynamical thermalization 
of the system is
simulated using standard Monte Carlo techniques.

We fix the initial spin configuration by mimicking the signs of  
a given vibrational mode. Starting with this spin configuration, 
we flip randomly $r$ spins and let the
system evolve towards the equilibrium situation assuming a noiseless scenario.
If the system recovers the initial configuration, the configuration is stable
under the flip of $r$ spins. The number of initial spin flips
determines the initial overlap, defined as $m_i=(N-r)/N$. After
dynamical evolution, the final overlap is given by $m_j=(N-s)/N$
where $s$ is the number of spins that differ from the initial
configuration. We repeat this process over $M$ initial different
random $r$ spin flips to evaluate statistically the final overlap
with the initial configuration: $m_f=(\sum_{j=1}^N m_j n_j)/M$,
$n_j$ being the number of times that the system reaches the $m_j$
configuration. The value of the initial overlap for which
significant decrease of the final overlap occurs, is a good measure
of the size of the basin of attraction of the corresponding pattern.
\begin{figure}
\includegraphics[width=0.9\linewidth]{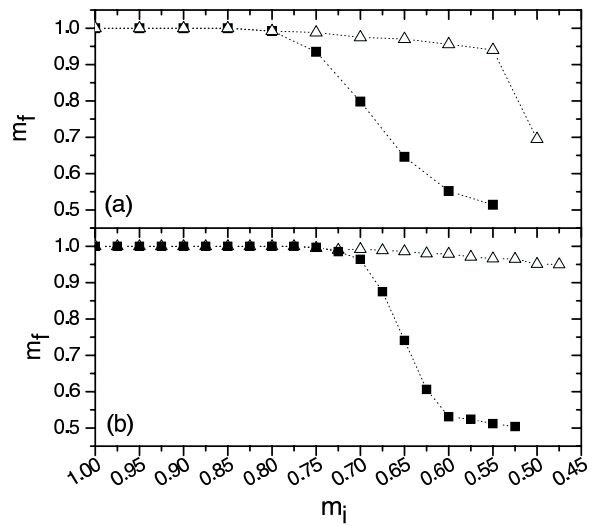}
\caption{Basin of attraction of the two lowest energy modes 
of an harmonically trapped ($\Omega_z=1 MHz$) chain of ions in the presence
of an external longitudinal magnetic field. 
(a) 20 Ca$^+$ ions with $ B= 7.7 \times 10^{-31}T m^2 C/s$ parallel to each
spin of the second mode and 
(b) 40 Ca$^+$ ions with $ B= 6.5 \times 10^{-31}T m^2 C/s$ parallel to each
spin of the second mode. Black squares (triangles) correspond to the lowest (second) mode.}
\label{fig_harmonic}
\end{figure}
For the harmonic trapping potential, the two lowest vibrational modes are the
center of mass (all spins parallel, with frequency $\omega_1=\Omega_z$) 
and the breathing mode 
(half up, half down, with $\omega_2=\sqrt{3} \Omega_z$) \cite{james}.
We find that the center of mass mode is stable up to the flip of
half of the spins, while the breathing mode is already unstable under
the flip of a single spin. Therefore, the only 
configuration that can be recovered is the center of mass mode, 
corresponding to all spins aligned up or down. To increase the
storage capacity of the network one can engineer the appropriate
spatial dependence of a longitudinal magnetic field to stabilize one
additional given pattern. We apply this technique to make the second
pattern stable. The results are displayed in Fig.\ref{fig_harmonic}
showing the basin of attraction of the two first modes for a chain
of (a) $20$ Ca$^+$ ions with $B=7.7 \times
10^{-31}T m^2 C/s $ parallel to each spin of the second mode and (b) 
$40$ ions for $B=6.5 \times 10^{-31}T m^2 C/s $ parallel
to each spin of the second mode. The results are averaged over 500
configurations and with random spin flip and random update. By
increasing (diminishing) the external magnetic field the stability
of the first pattern decreases (increases), while the second pattern
exhibits the complementary behavior. 
In  Fig.\ref{fig_harmonic}(a) we show that with
20 ions, it is possible to stabilize completely the first and second
pattern simultaneously up to the spin flip of 3 ions while the
probability of recovering the initial pattern for four initial
random spin flips is $99.2\%$ for both patterns. Increasing the
number of ions to 40, Fig.\ref{fig_harmonic}(b), the basin of
attraction increases up to the random flip of 9 ions with $100\%$ of
probability of recovering of the initial patterns, $99.7\%$
($99.5\%$) of probability of recovering the first (second) pattern
for the initial random flip of 10 spins and $98.5\%$ ($99\%$) of
probability of recovering the first (second) pattern for the initial
random flip of 11 spins.
\begin{figure}
\includegraphics[width=1.0\linewidth]{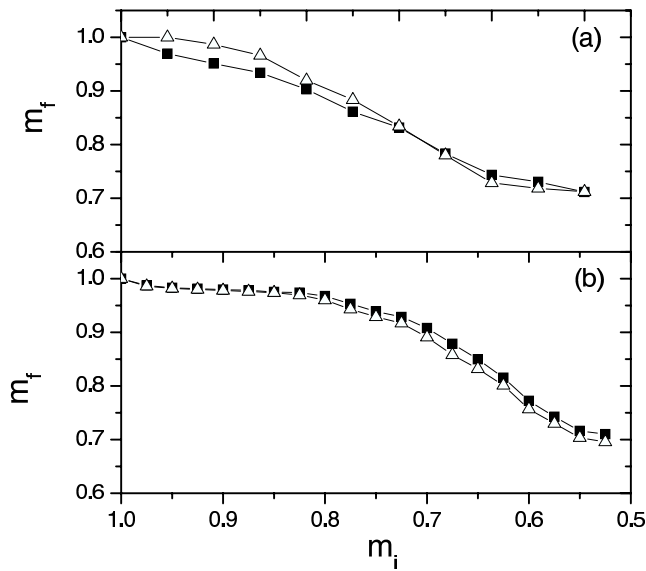}
\caption{Final overlap averaged over 500 initial configurations as a function of the initial overlap, for the two lowest vibrational modes of (a) 20 Ca$^+$ ions and (b) 40 Ca$^+$ ions in a potential $V=A|x|^{0.5}$ with $A=6.6\times 10^{-20}J/m^{1/2}$. The black squares (triangles)
correspond to the first (second) pattern.}
\label{FIG_POT}
\end{figure}

The second approach used to increase the storage
capacity of the network consists in modifying the spatial
dependence of the trapping potential. We consider an external
potential of the form $V(x)=A \mid x \mid^{\gamma}$ with
an even number of ions in order to avoid any possible singularity 
of the potential at the origin.
Fig.\ref{om1om2b} shows the ratio between the frequencies of the
second and the first mode $\omega_2/\omega_1$ as a function of
$\gamma$, for the case of $20$ ions. For $N\geq 20$ this ration does
not depend on the number of ions neither on the value of $A$. For
$\gamma =1$, and assuming small oscillations around the equilibrium
positions, the ratio $\omega_2/\omega_1$ is not well defined since
the second mode does not exist because of the constant character of
the force produced by the external potential. For $\gamma > 2$
severe limitations in the storage capacity of the network, similar
to those encountered in the harmonic case, appear. The situation
changes when $\gamma < 1$. In this case the two lowest modes become
nearly degenerated. We obtain larger number of stored patterns 
for $\gamma=0.5$ but the desired degeneracy is only obtained for the
first two collective modes. 
The results obtained with this optimal trapping potential are
displayed in Fig.\ref{FIG_POT} where the final overlap as a
function of the initial overlap for the two lowest vibrational modes
is depicted for (a) 20 Calcium ions and (b) 40 ions. 
The final overlap of the two lowest patterns and the
corresponding ones with all the spins flipped is close to 1 up to
two spin flips for $N=20$. This shows that the system is able to
recover the four patterns up to two spin flips. For
$N=40$ this holds up to eight spin flips. Note that the system does
not always recover the initial pattern but instead sometimes reaches
a slightly deformed configuration, which differs only in one spin flip from
the original one. For this reason $m_f$ in Fig. \ref{FIG_POT}(a) is
not exactly $1$ for one and two spin flips and up to eight in Fig.
\ref{FIG_POT}(b). In Fig. \ref{FIG_POT}(a) the probability of
recovering the initial pattern or one that differs in one spin flip
with the initial one is $100\%$ ($96.9\%$) for the second (first)
mode for the initial random spin flip of one spin and $98.7\%$
($95.1\%$) for the second (first) mode for the initial random spin
flip of two spins. In Fig. \ref{FIG_POT}(b) the probability of
recovering of the two modes is above $98\%$ up to three initial
random spin flips and above $97\%$ up to eight initial spin flips.

Summarizing, we have shown that spin-ion systems can be used to implement
NN models. We have calculated the storage capacity and robustness 
of such systems and showed how to control this storage capacity by changing 
the trapping potential and/or applying external longitudinal magnetic fields.
Furthermore, such models should allow to implement quantum neural networks 
by applying transverse magnetic fields that permit tunneling
processes between stored patterns. Finally, NN provide a clear example of
robust distributed information which could be exploited as 
a quantum memory.

We thank  I. Bloch, A. Bramon, H.-P. B\"uchler, G. Morigi, U. Sen, A. Sen (De), 
J. Wehr and P. Zoller for fruitful discussions.
We acknowledge support from the Deutsche
Forschungsgemeinschaft (SFB 407, SPP1078 and SPP1116), 
the RTN Cold Quantum Gases,
ESF PESC BEC2000+, the Spanish Ministerio de Ciencia y Tecnolog\'ia (MAT2002-
00699, FIS2005-04627), and the Alexander von Humboldt Foundation. V.A. acknowledges financial 
support from the Spanish Ministerio de Ciencia y Tecnolog\'ia.


\begin{references}
\bibitem{noteta}In \cite{porras}, the effective magnetic field $B'$ in Eq.
(\ref{hamiltonian}) includes a contribution from the pushing forces
caused by the applied lasers:
$B'=B+\frac{F^2}{m \Omega_z^2}$
This extra term due to the force is compensated by a constant
external magnetic field. Notice also that $F$ is such that it satisfies 
the validity of the derivation of the effective Hamiltonian: $\eta^2=\frac{F^2}{2 m \hbar \Omega_z^3}$ with $\eta=0.1$.
\end{references}
\end{document}